\documentclass[copyright,creativecommons]{eptcs}
\providecommand{\event}{WPTE 2017}

\providecommand{\firstpage}{1} 

\usepackage{tikz}
\usepackage{amsmath,amssymb,mathrsfs} 
\usepackage{graphicx}
\usepackage{url}
\usepackage{hyperref}
\usepackage[utf8]{inputenc}
\usepackage{rotating}
\usepackage{subfigure}
\usepackage{multirow}\usepackage{titlesec}
\usepackage{listings}
\usepackage{color}
\usepackage{cite}
\usepackage[ruled,vlined,linesnumbered,resetcount]{algorithm2e}
\usepackage{xspace}
\newcommand{\porgy}{{\sc Porgy}\xspace}

\SetAlgorithmName{Strategy}{strategy}{List of Strategies}
\newenvironment{ahpalgorithm}[1][htb]
 {\renewcommand{\algorithmcfname}{Algorithm}
  \begin{algorithm}[#1]
 }{\end{algorithm}}

\definecolor{darkgray}{rgb}{0.9,0.9,0.9}
\AtBeginDocument{}

\newcommand{\Ra}{\Rightarrow}
\newcommand{\ra}{\rightarrow}
\newcommand{\Interface}{\mathit{Interface}}
\newcommand{\X}{{\cal X}} 

\newtheorem{definition}{Definition}

\newtheorem{property}{Property}

\newenvironment{proof}{\paragraph{Proof.}}{\hfill$\square$}

\title {Attributed Hierarchical Port Graphs and Applications}
\author  {Nneka Chinelo Ene\institute{King's College London\\ Dept. of Informatics}\email{nneka.ene@kcl.ac.uk} \and Maribel Fern\'andez\institute{King's College London\\Dept. of Informatics}\email{maribel.fernandez@kcl.ac.uk} \and Bruno Pinaud \institute{University of Bordeaux\\ LaBRI UMR CNRS 5800, France}
 \email{bruno.pinaud@u-bordeaux.fr}}

\def\authorrunning{N. Ene, M. Fern\'andez, \& B. Pinaud}
\def\titlerunning{Attributed Hierarchical Port Graphs}

\begin{document}
\maketitle

\begin{abstract}
We present attributed hierarchical port graphs (AHP) as an extension of port graphs that aims at facilitating the design of modular port graph models for complex systems. AHP consist of a number of interconnected layers, where each layer defines a port graph whose nodes may link to layers further down the hierarchy; attributes are used to store user-defined data as well as visualisation and run-time system parameters. 
We also generalise the notion of strategic port graph rewriting (a particular kind of graph transformation system, where port graph rewriting rules are controlled by user-defined strategies) to deal with AHP following the Single Push-out approach.  We outline examples of application in two areas: functional programming and financial modelling. 
\end{abstract}

\section{Introduction}

We present a hierarchical graph transformation system built on a new formalism, that of an \emph{attributed hierarchical port-graph} (AHP).
Port graphs~\cite{Porgy_AndreiPhd} are graphs where edges are connected to nodes via ports. Attributed port graphs, where ports, nodes and edges carry data in the form of pairs attribute-value (i.e., records), have been used as a visual modelling tool in a variety of domains~\cite{porgy_2011,ValletKPM15,RuleBender}. They provide a visual mechanism to describe the structure of the system under study (via a graph) together with data structuring tools (attributes represent data associated to the model, as well as visualisation and run-time system parameters). However, whereas attributed port graphs are flat structures,  AHP are hierarchical structures that permit the nesting of graphs: nodes may have ``ladders'' to other graphs down the hierarchy. From the theoretical point of view, ladders are defined as additional structural components in the graph.  From a practical point of view, this can be implemented by using attributes of \emph{graph type} in records associated with nodes. In this sense, AHP could be thought of as a ``higher-order'' extension of attributed port graphs. The occurrence of attributes of graph type in nodes defines a forest structure (a hierarchy, where the graphs at level 0 are standard port graphs, and the nodes of graphs at level $i+1$ can include a ladder to a graph at level $i$). 
Formally, a hierarchical port graph is defined inductively using standard port graphs as base case.  
The restriction to a hierarchy of graphs (as opposed to arbitrary "graphs of graphs") simplifies the definition of AHP graph morphisms, which are a key notion in graph rewriting, and enforces a more natural and usable structure.

AHP forms the basis upon which graph transformation rules can be built that will enable a step-wise transformation of the graph. Unlike a regular port-graph transformation system, AHP rules fully capture hierarchical complexity without compromising on transparency and good visualisation. We propose the use of AHP to represent system states, and AHP rewrite rules to model the dynamic behaviour of the system. The rewriting relation generated by AHP rewrite rules can be controlled using the strategy languages already available for port graph transformation systems (see, e.g.,  \cite{porgy_2011,ValletKPM15}) to specify rewriting positions, rule priorities, etc.

We relate our notion of hierarchical graph rewriting to the conventional rewriting of flat graphs by introducing a flattening operation, which recursively replaces each hierarchical node by its contents resulting in a standard port graph. We show that under some conditions every hierarchical graph rewriting step gives rise to a standard rewriting step on the flattened graphs by using the flattened rule.

Hierarchical graphs have been used to specify re-factoring transformations for object-oriented programs~\cite{DBLP:journals/entcs/EetveldeJ03}, to provide semantics for distributed and mobile software systems with dynamic reconfiguration~\cite{Engels2000} and in computational biology~\cite{2006q.bio.....4006M} to define multi-layered biochemical systems and represent molecular agents at varying levels of abstraction~\cite{grs_stochastic_kappa_only_refinement_rules_}, to name but a few areas. 
With the aim of improving the support for graph-based programming languages, hierarchical hyper-graph transformation systems have been previously proposed, where certain hyper-edges contain hyper-graphs~\cite{grs_hgraph_hierarchical_pl_2000}, and rewriting is defined following the double push-out approach. In this paper, we focus on port graphs, which follow the single push-out (SPO) approach to rewriting, and provide an SPO semantics for AHP. 

We illustrate the use of hierarchical port graphs in two areas: securitisation (specifically, we model the secondary market ``rational negligence phenomenon''~\cite{FIN_agents_absmodel_,fin_sec_model}, whereby investors may choose to trade securities without performing independent evaluations of the underlying assets) and functional programming (specifically, $\lambda$-term evaluation). 

Securitisation models have been widely studied especially after the 2008 crisis (see, e.g., \cite{bath48096,RePEc:pal:jbkreg:v:14:y:2013:i:3:p:285-305}) and a port graph model illustrating the rational negligence problem is available~\cite{WADT_Paper_}. In this paper we show how the hierarchical features of AHP can be used to refine and complete this model,  catering for the more operational tiers of the securitisation process. Whilst the top tier of the rational negligence model deals with asset transfers, the operational tiers, at a basic level, encapsulate asset origination, packaging, structuring and servicing details, the key processes in the life cycle of a structured security. 

Graph representations have been used in many $\lambda$-calculus evaluators to enforce sharing of computations; interaction nets (a particular kind of graph rewriting formalism introduced by Lafont~\cite{DBLP:conf/popl/Lafont90}, inspired by the graphical notation of linear logic proofs) permit to implement interesting strategies of evaluation, such as optimal reduction~\cite{DBLP:conf/popl/GonthierAL92}. However, in interaction net evaluators the representation of $\lambda$-abstraction is involved due to the fact that $\lambda$ is a binder and explicit markers have to be used to define and manage its scope. Higher-order port graphs~\cite{grs_portgraph_hierarchical} permit to encode binders in a direct way and can be easily represented as AHP. In future work we will explore the links between hierarchical and higher-order port graphs. 

 Summarising, our main contribution is a definition of attributed hierarchical port graph, together with corresponding notions of graph morphism and rewriting relation, following the SPO approach. To highlight the suitability of attributed hierarchical port graphs as a conceptual framework we give two examples of application and compare the obtained AHP models with closely related models already available (based on standard port graphs or higher-order port graphs). 
 
This paper is organised as follows: We recall key notions useful in our analyses in Sect.~\ref{section_background}. Attributed Hierarchical Port Graphs are defined in Sect.~\ref{section_hierarchicalportgraphs}. Examples are described in Sect.~\ref{section_gtsmodel}. Sect.~\ref{section_properties} examines key properties of AHP graphs. Sect.~\ref{section_related_work} discusses related work. We finally conclude and  briefly outline future plans in Sect.~\ref{section_conclusion}.

\section{Background}
\label{section_background}
 
There are many different kinds of graph transformation systems see, for instance, \cite{Plump98termgraph,Corradini:handbook,Plump98termgraph,HabelMP01}. In this paper we examine the transformation of port graphs~\cite{grs_porgy_main_long}, which have  been used as a modelling tool in various domains, such as biochemistry and social networks~\cite{RuleBender,ValletKPM15}.

Intuitively an attributed port graph is a  graph where nodes have explicit connection points, called ports, and edges are attached to ports. Nodes, ports and edges are labelled by a set of attributes describing  properties such as colour, shape, etc. To formally define attributed port graphs, we follow~\cite{grs_porgy_main_long}, where records (i.e., sets of pairs attribute-value) are attached to graph elements. 

\begin{definition}[Signature]
\label{def:sig}
A port graph signature $\nabla$ consists of   the following pairwise disjoint sets:
$\nabla_\mathscr{A}$, a set of attributes; $\X_\mathscr{A}$, a set of attribute variables;
$\nabla_\mathscr{V}$, a set of values; $\X_\mathscr{V}$, a set of value variables.
\end{definition}

\begin{definition}[Record] 
 A record  $r$ over the signature $\nabla$ is a set
 $\{(a_1,e_1),\ldots, (a_n,e_n)\}$ of pairs, where  for $1 \leq i\leq n$, $a_i \in
 \nabla_\mathscr{A}\cup \X_\mathscr{A}$ and $e_i$ is an expression built from  
 $\nabla_\mathcal{A}\cup \nabla_\mathscr{V}\cup \X_\mathscr{V}$, each
 $a_i$ occurs only once as first component of a pair in $r$, and 
 there is one pair where $a_i =
 Name$, a special element. The function $Atts$ applies to records and returns
 the labels of all the attributes: 
$Atts(r)= \{a_1,\ldots,a_n\}$ if $r = \{(a_1,v_1),\ldots, (a_n,v_n)\}$.
As usual, $r.a_i$ denotes the value $v_i$ of the attribute $a_i$ in $r$.
The attribute $Name$ identifies the record in the following sense: $\forall r_1,r_2, Atts(r_1)=Atts(r_2)$ if $r_1.Name = r_2.Name$. 
\end{definition}

In addition to $Name$, records may contain any number of data attributes, which must be of basic type (i.e., numbers, strings, Booleans, etc.).

\begin{definition}[Port Graph] 
\label{def:portgraphs} 
A {\em port graph} over a {\em signature}
$\nabla$ is a tuple $G=( V,P, E,D)_{\cal F}$ where
$V$ is a finite set of nodes ($n, n_1, \ldots$ range over nodes);
$P$ is a finite set of ports ($p, p_1, \ldots$ range over ports); 
$E$ is a finite set of edges between ports ($e,
  e_1, \ldots$ range over edges; edges are undirected and two ports may be connected 
  by more than one edge);
$D$ is a set of records over $\nabla$, and 
${\cal F}$ is a set of functions 
$Connect\colon E \ra P \times P$, $Attach\colon P \ra V$ and ${\cal L}\colon V \cup P\cup E \ra D$ such that
\begin{itemize}
\item for each edge $e\in E$, $Connect(e)$ is the pair $ \{p_1,p_2\}$ of ports connected by $e$ (an ordered pair if the edge is oriented);
 \item for each port $p\in P$, $Attach(p)$  is the node to which $p$ belongs.
 \item ${\cal L}$ is a labelling function that returns a record for each element in $V \cup P\cup E$, such that for each node $n \in V$, ${\cal L}(n)$ contains an attribute $\Interface$ whose value is the list of names of the ports attached to $n$, that is, ${\cal L}(n).\Interface = [{\cal L}(p_i).Name \mid Attach(p_i) = n]$, satisfying the following constraint:
${\cal L}(n_1).Name  = {\cal L}(n_2).Name \Rightarrow {\cal L}(n_1).\Interface = 
{\cal L}(n_2).\Interface.$ 
\end{itemize}
Similarly, we call \emph{Interface} of a graph its set of free ports (i.e., ports that do not have edges attached to them).
\end{definition}
The functions $Connect$ and $Attach$ can be implemented as attributes in the records associated to edges and ports, respectively.
 
A port graph rewriting rule $L \Rightarrow_C  R$ can itself be seen as a port graph consisting of two port graphs $L$ and $R$ together with an ``arrow'' node linked to $L$ and $R$ by a set of edges that specify a mapping between ports in $L$ and $R$ (see Figure~\ref{figure:updateladder2} for an example of a rule, the edges involving the arrow node are red in the figure). The pattern, $L$, is used to identify sub-graphs in a given graph which should be replaced by an instance of the right-hand side, $R$, provided the condition $C$ holds. The edges involving the arrow node indicate how the instance of $R$ should be linked to the remaining part of the graph in a rewriting step (they specify the rule morphism in the SPO approach). Each of the ports attached to the arrow node has an attribute Type $\in \nabla_{\mathcal{A}}$, which can have three different values: bridge, wire and black-hole. A port of type bridge must have edges connecting it to $L$ and to $R$ (one edge to $L$
and one or more to $R$); 
a port of type black-hole must have edges connecting it only to $L$ (at least one edge);  
a port of type wire must have exactly two edges connecting to $L$ and no edge connecting
to $R$.	

Intuitively, a port of type bridge in the arrow node connecting to $p_1$ in $L$ and $p_2$ in $R$ indicates that $p_1$ survives the reduction and becomes $p_2$. A port in L connected to a black-hole port in the arrow node does not survive the reduction; all edges connected to this port in the graph are deleted when
the reduction step takes place. A port of type wire connected to two ports $p_1$ and $p_2$ in the left-hand side triggers a particular rewiring, which takes all the
ports that are connected to (the image of) $p_1$ in the redex and creates an edge for each of those ports to each of the ports connected to (the image of) $p_2$. We refer to~\cite{grs_porgy_main_long} for more details and examples.

To define the rewrite relation, we use port graph morphisms, which preserve the graph structure and the values of the attributes (instantiating variables that occur in patterns).

\begin{definition}[Port-graph Morphism\!\hspace*{1pt}]
 Let $G = (V_G,P_G, E_G,D_G)_{{\cal F}_G}$ and 
 $H= (V_H,P_H, E_H,D_H)_{{\cal F}_H}$ be port graphs over the same signature $\nabla$, 
 a {\em morphism} $f$  from $G$ to $H$, denoted $f: G \rightarrow H$, is a family of (partial) functions $\langle f_V:V_G \ra V_H, f_P: P_G \ra P_H, f_E:
 E_G \ra E_H, f_D\colon D_G \ra D_H \rangle$ 
 such that 
 \begin{itemize}
 \item $f_V, f_P, f_E$ are injective, i.e., the morphism does not identify distinct nodes, ports or edges;
\item $\forall e \in E_G$, if  $Connect_G(e) = \{p_1,p_2\}$ then $\{f_P(p_1), f_P(p_2)\} = Connect_H(f_E(e))$, i.e., the morphism preserves the edge connections;
\item $\forall n \in V_G$, if $Attach_G(p)=n$ for some $p$ then $f_{V}(n) = Attach_H(f_P(p))$, i.e., the morphism preserves the port attachments;
\item For all $n \in Dom(f)$,
$f_{D}( {\cal L}_G(n) ) = {\cal L}_H (f_V(n))$  \\
For all $p \in Dom(f)$,
$f_{D}( {\cal L}_G(p) )= {\cal L}_H (f_P(p))$ \\
For all $e \in Dom(f)$,
$f_{D}( {\cal L}_G(e) ) = {\cal L}_H (f_E(e))$,  \\
i.e., the morphism preserves attributes and their values; note that $f_D$ may instantiate variables.
\end{itemize} 
We denote by $f(G)$ the subgraph of $H$ consisting of the set of nodes, ports, edges and records that are images of nodes, ports, edges and records in $G$.
\label{def:morphism}
\end{definition}

This definition ensures that each corresponding pair of nodes, ports and edges in $G$ and $H$ have the same set of attribute labels and associated values, except at positions where there are variables. When using this definition to define rewriting, the left-hand
side of the rewrite rule may include variable labels but the graph to be rewritten will not have variables.

\begin{definition}[Match]
\label{def:match}
Let $L \Rightarrow_C R$ be a port graph rewrite rule and $G$ a port graph. We say a match $g(L)$ of $L$ (i.e., a redex) is found in $G$ if there is a port graph morphism $g$ from $L$ to $G$ (hence $g(L)$ is a sub-graph of $G$), $C$ holds in $g(L)$, and for each port in $L$ that is not connected to the arrow node, 
its corresponding port in $g(L)$ is not an extremity in the set of edges of $G - g(L)$. 
\end{definition}
 
Graph rewriting systems can be given a categorical semantics: the most popular approaches are the single pushout and double pushout semantics~\cite{DBLP:conf/gg/EhrigHKLRWC97}. In this paper we are interested in the category $\mathsf{C}$  whose objects are attributed hierarchical port graphs  and whose morphisms are hierarchical morphisms (defined in the next section). We recall now the notion of a pushout from~\cite{grs_hgraph_hierarchical_pl_2000}, which is used to define rewriting steps.  

\begin{definition}[Pushout]
A pushout in category $\mathsf{C}$ is a tuple $(a_1,a_2,b_1,b_2)$ of morphisms $a_i: X \rightarrow X_i$, $b_i: X_i \rightarrow X'$ such that $b_1 \circ a_1 = b_2 \circ a_2$ and for all $b'_i: X_i \rightarrow C$ where $i=\lbrace1|2\rbrace$ such that $b'_2 \circ a_2 = b'_1 \circ a_1$ there exists a unique morphism $c: X' \rightarrow C$ that satisfies $c \circ b_1=b'_1$ and $c \circ b_2=b'_2$. 
\end{definition}

Let $G$ be a port graph.  A \emph{rewrite step}  $G \Rightarrow H$ via the port graph rewrite rule $L \Rightarrow_C R$ is obtained by replacing in $G$ a match $g(L)$  by $g(R)$ and redirecting the edges incident to ports in $g(L)$ to ports in $g(R)$ as indicated by the arrow node. 
The last point in Definition~\ref{def:match} ensures that ports in $L$ that are not connected to the arrow node are mapped to ports in $g(L)$ that have no edges connecting them with ports outside the redex, thus ensuring that there will be no dangling edges when $g(L)$ is replaced by $g(R)$. Under suitable conditions (see \cite{grs_porgy_main_long} for details), the rewriting relation can be given a semantics using the SPO construction: a rewriting step is obtained by computing the pushout of the rule morphism (defined by the arrow node and its edges) and the matching morphism. 
$\mathcal{S}$ can be applied to $x$ we say that the strategy has failed.

\porgy~\cite{grs_porgy_main_long} is an interactive environment that includes functionality to create port graphs and port graph rewrite rules, and to apply rules to graphs (according to user-defined strategies). In this implementation, the functions $Connect$ and $Attach$ (see Definition~\ref{def:portgraphs}) are represented as attributes in records (i.e., records contain data attributes, visualisation attributes such as colour or shape, and structural attributes such as Connect and Attach). \porgy also  provides a visual representation of the rewriting derivations, which can be used to analyse the rewriting system; however, it lacks mechanisms to define graphs in a modular way. To facilitate the design of modular port graph models, we plan to extend this tool with attributed hierarchical graphs (presented in the next section). 

\section{Attributed Hierarchical Port-graphs}
\label{section_hierarchicalportgraphs}
We extend the notion of a port graph  to support a multi-level structure. 
The key idea is the introduction of a new function on nodes, called $Ladder$, which returns a graph. This function will be implemented as a new kind of attribute in records associated with nodes, which we also call \emph{Ladder} and whose value is of type graph. 
In the definition of AHP we use the concept of \emph{Interface} of a graph, which is the set of free ports in the graph (i.e., ports that do not have edges attached to them).

The set of  attributed hierarchical port graphs will be defined by induction. More precisely, AHP  will be defined by levels, where the graphs at level $i$ may use graphs of a lower level only.  
 The levels induce a notion of hierarchy in the graph: in an AHP, the nesting of graphs defines  a tree structure. This reduces the complexity of the matching problem, ensures compositionality (which is 
 important when defining rewriting and flattening relations: components of the graph can be replaced by other graphs and the resulting overall graph remains well formed; it is more difficult  to ensure that no edges are left dangling if there are edges across components), and facilitates the visualisation of the graph, generating a greater ease in user-understanding. More flexible notions of hierarchical graphs will be considered in future work.
 
\begin{definition}[AHP-Signature]
An AHP-signature $\nabla_\mathcal{G}$ consists of a port graph signature (see Definition~\ref{def:sig}), where the set of variables includes elements of type graph.
The subset $\mathcal{X}_\mathcal{V}^G$ of $\mathcal{X}_\mathcal{V}$ consists of variables of type graph, denoted $\mathfrak{X}_1, \mathfrak{X}_2, \ldots$, representing unknown port graphs, each with an associated interface (list of port names) denoted by $\Interface(\mathfrak{X})$. 
\end{definition}
 
\begin{definition}[Attributed Hierarchical Port-graph] 
\label{def:AHP}
An AHP (\emph{Attributed Hierarchical Port-graph}) over a signature $\nabla_\mathcal{G}$ is an element of the set
$\mathcal{H} = \bigcup_{i \geq 0} \mathcal{H}_i$, where $\mathcal{H}_i$, the set of AHP at \emph{level} $i$, is defined as follows:

An AHP at level 0 is an attributed port graph as specified in Definition~\ref{def:portgraphs}
or a variable $\mathfrak{X}$ of type graph. 

An AHP at level $i+1$ is a tuple 
$(V, P, E, {\cal G},D)_\mathcal{F}$ consisting of a set $V$ of nodes, a set $P$ of ports,  a set $E$ of edges, a set ${\cal G}\subseteq \bigcup_{j \leq i} \mathcal{H}_j$ of AHP at a lower level, and a set  $D$ of records over $\nabla_\mathcal{G}$, together with a set of functions $Connect$, $Attach$, $Ladder$ and $\mathcal{L}$ such that $Connect$ and $Attach$ are defined as in Definition~\ref{def:portgraphs}, $Ladder\colon V \mapsto {\cal G}$ is a partial injective function mapping nodes to lower-level graphs that must have the same interface as the node (i.e.,  the Ladder graph should have same number of free ports, with the same names and attributes, as the node), and  $\mathcal{L}$ is a labelling function that associates records in $D$ to elements in $V,P,E, {\cal G}$. Given an AHP $G$, we assume that the Ladder graphs  within $G$ are pairwise disjoint and also disjoint with the top level graph (that is, they do not share nodes, ports or edges, and therefore there are no edges across graphs at different levels in $G$, or graphs of the same level in different nodes).
\end{definition}

Figure~\ref{figure:starting1} shows an example of an AHP graph: the node $A$ in the top level has a Ladder whose value is the  graph shown in the centre of the figure, which in turn contains a node $Pools$ with a Ladder graph (the graph in the rightmost part of the figure). 

To define the rewriting relation, we need AHP-morphisms. The definition of AHP-morphism  (Definition~\ref{def:AHPmorphism}) ensures that corresponding data attributes in $G$ and $H$ have the same values (except at positions where there are variables in $G$, which are instantiated in $H$), and attachment of ports, edge connections and ladder graphs are preserved. If $f$ is a morphism from $G$ to $H$, we will denote by $f(G)$ the sub-graph of $H$ consisting of the set of nodes, ports, edges, ladder graphs and records that are images of nodes, ports, edges, ladders and records in $G$.

\begin{definition}[AHP Morphism]
\label{def:AHPmorphism}
AHP-morphisms are defined inductively. 

An AHP at level 0 is either an attributed  port graph or a variable of type graph, consequently an AHP-morphism at level 0 is a port graph morphism or a mapping from graph variables to AHP.

Let $G = (V_G,P_G, E_G,{\cal G}_G, D_G)_{{\cal F}_G}$  and 
$H= (V_H,P_H, E_H,{\cal G}_H,D_H)_{{\cal F}_H}$ be two AHP graphs over the same signature $\nabla_{\mathcal{G}}$; assume w.l.g.\ that $G$ is at level $i$. 
 A (partial) {\em AHP morphism at level $i$}, $f$,  from $G$ to $H$, denoted $f: G \rightarrow H$, with definition domain $Dom(f)$, is defined by a family of
(partial) functions $\langle f_V:V_G \ra V_H, f_P: P_G \ra P_H, f_E:
E_G \ra E_H, f_{{\cal G}}: {\cal G}_G \mapsto {\cal G}_H, f_D: D_G \mapsto D_H\rangle$ such that
\begin{itemize}
\item $f_V, f_P, f_E, f_{\cal G}$ are injective.
\item For all $e \in E$, if $Connect_G(e) = \{p_1,p_2\}$ then $Connect_H(f_E(e)) = \{f_P(p_1), f_P(p_2)\}$ (i.e., the morphism preserves the edge connections).
\item For all $n \in  V_G$, if $Attach_G(p) = n$ for some $p$ then $Attach_H(f_P(p))= f_V(n)$
(i.e., the morphism preserves the attachment of ports to nodes), and if $Ladder_G(n) = W$ then $Ladder_H(f_V(n)) = f_{\cal G}(W) $, where $f_{\cal G}$ is an AHP morphism at level $j < i$ (i.e., the morphism preserves the hierarchical structure).
\item 
For all $n \in Dom(f)$,
$f_{D}( {\cal L}_G(n) ) = {\cal L}_H (f_V(n))$  \\
For all $p \in Dom(f)$,
$f_{D}( {\cal L}_G(p) )= {\cal L}_H (f_P(p))$ \\
For all $e \in Dom(f)$,
$f_{D}( {\cal L}_G(e) ) = {\cal L}_H (f_E(e))$  \\
For all $W \in Dom(f)$,
$f_{D}( {\cal L}_G(W) ) = {\cal L}_H (f_{\cal G}(W))$  \\
This constraint ensures that the morphism preserves record
attributes and their values;
note that $f_D$ may instantiate variables.
\end{itemize}
\end{definition}

As in the case of standard port graphs, when using morphisms to define rewriting, we will only allow the use of variable labels on one of the graphs: the graph $L$ on the left-hand side may include variable labels, whilst the graph to be rewritten may not. 

\begin{definition}[AHP Rewrite Rule]
\label{def:AHPrule}
An AHP rewrite rule $L \Rightarrow_C R$ is an AHP graph that consists of two sub-graphs $L$ and $R$ together with a node (called arrow node) that may carry a condition $C$ and whose edges link to ports in the top level of $L$ and  in the top level of $R$, capturing the correspondence between top level ports in $L$ and $R$. The three types of ports in the arrow node (i.e. bridge, wire and black-hole) remain the same as for conventional port graphs.
\end{definition}
Figure~\ref{figure:updateladder1} is an example of an AHP rule. 

\begin{definition}[AHP Match]
\label{def:AHPmatch}
Let $L \Rightarrow_C R$ be an AHP rewrite rule and $G$ an AHP graph.
A match between $L$ and $G$ is said to take place if a total AHP morphism $g$ between $L$ and a sub-graph of $G$ can be identified, such that $g(L)$ satisfies $C$ and for each port in $L$ that is not connected to the arrow node, its corresponding port in $g(L)$ is not an extremity in the set of edges of $G - g(L)$.    
\end{definition}

Note that to find a match between $L$ and $G$, if a node in $L$ has a ladder graph $W$ then a new morphism is sought (recursively, following the inductive definition of AHP morphism) between  $W$ and the ladder graph in the corresponding node in $G$.

The \emph{rewriting steps generated by AHP rules}, denoted $G \Rightarrow H$,  are defined in a similar way as for attributed port graphs (see Section~\ref{section_background}): a subgraph $g(L)$ in $G$ is replaced by $g(R)$ and the edges incident to $g(L)$ are rewired as indicated by the arrow node, provided the morphism $g$ satisfies the matching conditions.
We study properties of the rewriting relation and provide a SPO semantics in Section~\ref{section_properties}, after giving examples of applications in the next section.

\section{Applications} 
\label{section_gtsmodel}
\paragraph{Securitisation.}
As defined in \cite{FIN_ABM_Essex_CRTransferModel_} ``Securitisation is the process of converting cash flows arising from underlying assets or debts/receivables (typically illiquid such as corporate loans, mortgages, car loans and credit cards receivables) due to the originator into a smoothed liquid marketable repayment stream" and this ensures that the originator can raise asset-backed finance through loans or the issuance of debt securities. An \emph{originator} is any financial intermediary with a portfolio of assets on its balance sheet.  \emph{Assets} represent loans to clients or \emph{obligors} who make regular installment payments to the originator to clear their debts. In a securitisation, assets are selected, pooled and transferred to a tax neutral, bankruptcy-remote, liquidation-efficient (i.e bankruptcy avoiding), \emph{special purpose entity} (SPE) or \emph{special purpose vehicle} (SPV), who funds them by issuing \emph{securities}. In general, an ABS (asset-backed security), or simply asset if there is no ambiguity, is any securitisation issue backed by consumer loans, car loans, credit cards, etc. The ABS trading model described in~\cite{FIN_agents_absmodel_} has been specified as a port graph rewriting system and implemented in \porgy~\cite{WADT_Paper_}; in this paper we incorporate lower, more operational tiers into the model, by means of AHP. The operational tiers will lead to the computation of the asset's ``pay off'' attribute at any point in time (in this case, the profit made from successfully reselling the security, given varying toxicity likelihood values and the internal examination of the time-value-of-money).

\begin{figure}[h!] 
\begin{center}
        \includegraphics[scale=0.3815]{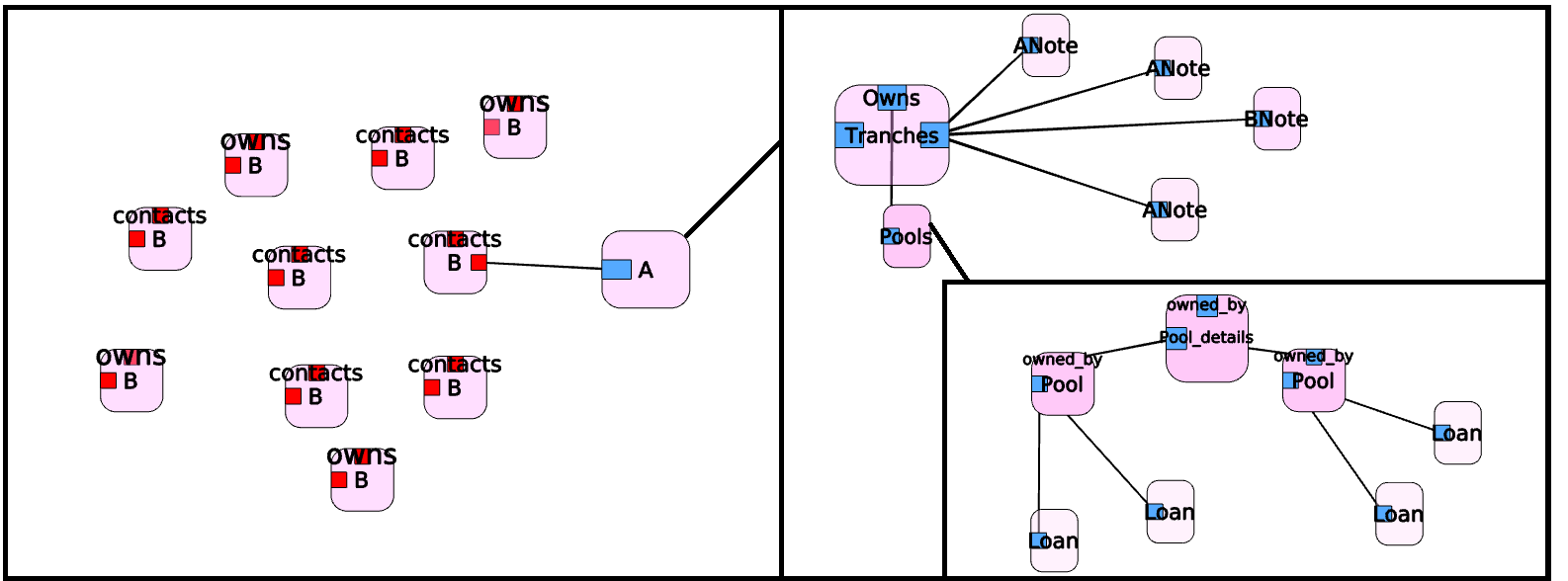}
        \caption{Sample Starting Graph}
        \label{figure:starting1}  
\end{center}
\end{figure}

We represent the full ABS universe hierarchically and enforce information flow bidirectionally. AHP rules and strategies control the step-wise evolution of the graphs. The derivation tree  is used for plotting and analysing parameters. The asset trading model~\cite{WADT_Paper_} sits at the top level of the model hierarchy and below this system lie several more deterministic subsystems that encapsulate asset origination, packaging, structuring and servicing processes, and therefore aid in enforcing internal checks as a result of complete system integration. Figure~\ref{figure:starting1} depicts the main components of the securitisation model. The first, second and third graphs respectively represent (simplified versions of) the Secondary Market, Structuring and Origination tiers respectively. B is an agent, the buyer or seller bank (there are ten nodes representing banks in the top level), node A is an asset-backed security, node Tranches represents the internal structuring or content of an asset and Pools the connection to underlying that provide an income stream.

\begin{figure}[h!]
    \begin{center}
        \includegraphics[scale=0.3]{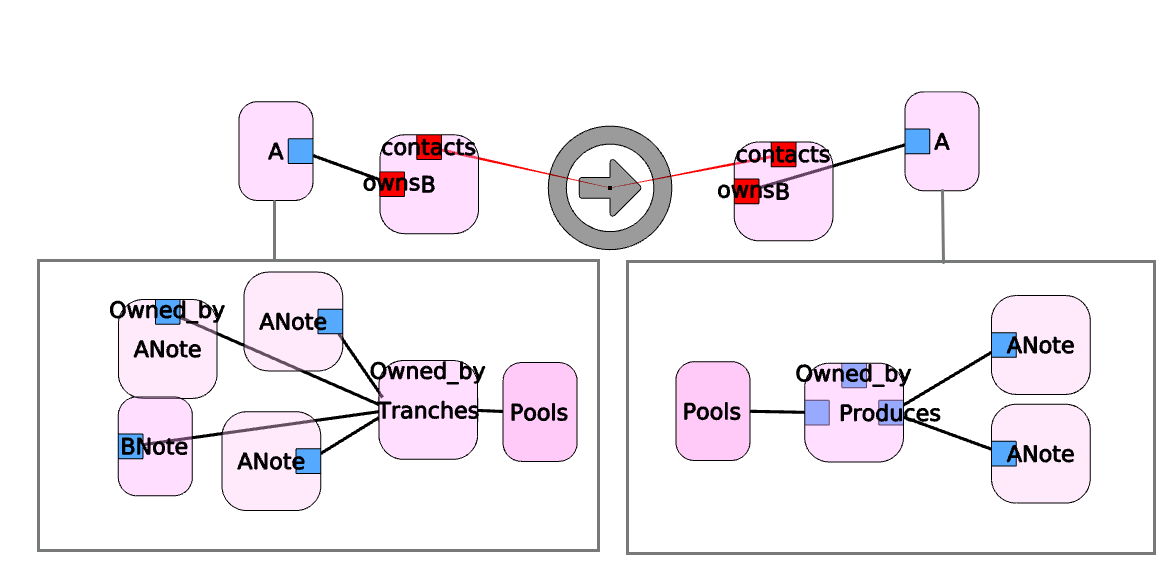}
        \caption{Sample Hierarchical Rewrite Rule: ``Update Ladder''}
        \label{figure:updateladder1}
    \end{center}
\end{figure}

\begin{figure}[h!]
    \begin{center}
        \includegraphics[scale=0.3]{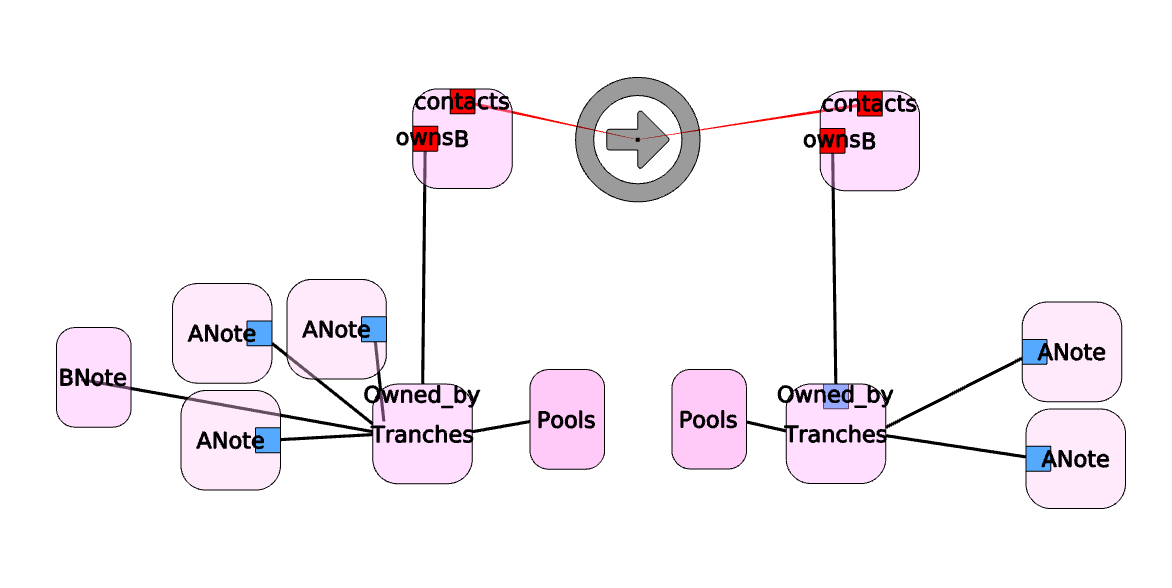}
        \caption{Flattened Version of Sample Hierarchical Rewrite Rule: ``Update Ladder''. The Ladder graph of node A is updated.}
        \label{figure:updateladder2}
    \end{center}
\end{figure}

The  hierarchical rewrite rules that drive execution have patterns that cut across the two or three identified tiers. The transformation of the context graph can be seen in asset transfers,
general node movements, and colour changes that indicate changes of state. Rules such as that which aid in the structuring of the asset as seen in Figure~\ref{figure:updateladder1} or its flattened version as seen in Figure~\ref{figure:updateladder2}, can be applied from within the system.

A flattening algorithm that transforms AHP into standard attributed port graphs as seen in the sample rule is described in the next section. 
Compared with standard port graph models, where all the different features and processes have to be represented in the same "flat" graph, the hierarchical model facilitates the analysis of the system as it is possible to hide the non-relevant details in lower levels to focus on the features of interest. 

\paragraph{Lambda-terms:}
\label{section_hopgahp}
The $\lambda$-calculus~\cite{DBLP:books/el/leeuwen90/Barendregt90} is a paradigmatic model of functional computation. We now consider common $\lambda$-term representations, and investigate  AHP encodings.

Intuitionistic logic proofs (which, by the Curry-Howard isomorphism are equivalent to $\lambda$-terms) expressed in a natural deduction style  can be inductively translated into conventional port graphs (standard interaction net translations can be used, since interaction nets are a particular kind of port graph).
 
This representation works well for some aspects of logic but not where boxes are required  as this introduces a two level structure that is not available in standard port graphs or interaction nets.
Boxes are represented by extra nodes requiring additional rules for book-keeping. To palliate this problem, several extensions of interaction nets have been proposed (see e.g.,~\cite{Accattoli15,FernandezMP07,grs_portgraph_hierarchical1}).
In~\cite{grs_portgraph_hierarchical1}, a representation of intuitionistic logic proofs (or $\lambda$-terms) is given by means of \emph{higher-order port-graphs} (HOPG).

HOPG~\cite{grs_portgraph_hierarchical} extend port graphs with \emph{higher-order variable nodes}, which can be instantiated by port graphs.  

A labelled higher-order port-graph consists of a collection of first-order nodes (whose names can be constants or variables), a collection of higher-order nodes (whose names are variables), a collection of ports attached to nodes (higher-order variable nodes contain only variable ports), undirected edges, and  labelling functions that determine the concrete properties (attributes and values) of each graph element.   

AHP graphs  subsume  HOPGs  by introducing an abstraction level that not only fulfils the original HOPG objective of simulating ``boxes'' or the grouping of a collection of nodes within one node, all interfaces matching, but that also maintains a nesting structure that can be recursively flattened on demand to produce a conventional graph. Similar to the example HOPG-implementation given in~\cite{grs_portgraph_hierarchical}, an AHP graph can directly  represent a proof (or a $\lambda$-term) that contains a box by  using a hierarchical node $Box$ whose Ladder  contains the box structure. 

In general,  to encode an HOPG as an AHP graph we simulate higher-order variable nodes by introducing an abstraction level. More precisely, a higher-order variable node labelled by $\mathfrak{X}$ in the HOPG is represented by a node with a Ladder with value $\mathfrak{X}$ in the corresponding AHP graph. In other words, where a higher-order variable is used in an HOPG to represent an unknown subgraph,  a node with a Ladder containing a graph variable is used. The parent node can be seen as a place-holder, named appropriately. It maintains an interface that the sub-graph, when instantiated, will also adopt. 

The encoding of HOPG rules is more involved; a detailed definition is left for future work.

\section{Properties}
\label{section_properties}

\paragraph{Soundness:}
 Rewriting an AHP produces another AHP, that is, rewriting does not leave dangling edges and maintains a hierarchical structure. 

\begin{property}
Let $G$ be an AHP and $L \Ra_C R$ an AHP rewrite rule.

If $G\Ra H$ using   $L \Ra_C R$  then $H$ is an AHP. 
\end{property}

\begin{proof}
We need to check that there are no dangling edges after rewriting, and the result is a hierarchical structure.

The restriction to \emph{injective} Ladder functions in the definition of AHP 
ensures that the matching morphism will not identify two ladder graphs (which would  break the hierarchy). Moreover,  since a rule is an AHP, the arrow node can only link ports in $L$ and $R$ at the top level (no edges can cross level boundaries) and therefore the replacement of the subgraph $g(L)$ by $g(R)$  maintains a hierarchical structure: the rewiring (during rewriting) cannot introduce edges across subgraphs at different level.  

A rewriting step cannot leave dangling edges due to the constraints imposed in the definition of rewriting,  similar to the constraints used in the definition of regular port graph rewriting: the rewiring phase takes care of the edges arriving from outside the redex to images of ports connected to the arrow node, and there are no other edges between ports outside the redex and ports in the redex (as specified in the definition of rewriting). Hence, there are no dangling edges when $g(L)$ is replaced with $g(R)$ in an AHP-rewriting step. 
\end{proof}

\paragraph{Flattening:}
\label{subsection_flattening}

\begin{ahpalgorithm}[t!]
Let $G= (V_G,P_G, E_G,{\cal G}_G, D_G)_{{\cal F}_G}$ be an AHP graph without variables of type graph, \\
let $V_G^0$  be the subset of $V_G$  where the function Ladder is not defined and  $P_G^0$ the set of ports attached to nodes in $V_G^0$.\\
$\mathcal{T}(G) = G$ ~~~~ if $V_G^0 = V_G$ (i.e., $G \in \mathcal{H}_0$), \\
$\mathcal{T}(G) = G'$ ~~~ otherwise, where $G'= (V_{G'},P_{G'}, E_{G'}, D_{G'})_{{\cal F}_{G'}}$ such that  \\
~~~~~~~~~~~~~~~~~~~~ ${\cal W} = \{ \mathcal{T}(Ladder(n)) \mid  n \in dom(Ladder_G)\}$  \\
~~~~~~~~~~~~~~~~~~~~ $V_{G'} = V_G^0 \cup \bigcup_{W \in {\cal W}} V_W$\\
~~~~~~~~~~~~~~~~~~~~ $P_{G'} = P_G^0 \cup \bigcup_{W \in {\cal W}} P_W$\\
~~~~~~~~~~~~~~~~~~~~ $E_{G'} = E_G \cup \bigcup_{W \in {\cal W}} E_W$\\
~~~~~~~~~~~~~~~~~~~~ $D_{G'} = D_G \cup \bigcup_{W \in {\cal W}} D_W$\\

~~~~~~~~~~~~~~~~~~~~ $Connect_{G'} =  \overline{Connect}_G \cup \bigcup_{W \in {\cal W}} Connect_W$\\
~~~~~~~~~~~~~~~~~~~~ $Attach_{G'} = Attach_G|_{P_{G'}} \cup \bigcup_{W \in {\cal W}} Attach_W$\\
~~~~~~~~~~~~~~~~~~~~ ${\cal L}_{G'} = {\cal L}_G|_{V_{G'}\cup P_{G'}\cup E_{G'}} \cup \bigcup_{W \in {\cal W}} {\cal L}_W$\\
\caption{Flattening Function $\mathcal{T}$}\label{alg:alg5}
\end{ahpalgorithm} 

In order to relate our notion of AHP rewriting to the conventional transformation of flat port graphs,  and to test our example model in the current version of \porgy, we have implemented a flattening function $\mathcal{T}$ that unfolds an AHP without graph-type variables into a regular port graph as seen in Algorithm~\ref{alg:alg5}. Since the  Ladder graph of  a node $n$ has the \emph{same interface} as $n$, we can flatten the graph $G$ by replacing each hierarchical node with a flattened version of its Ladder graph (recursively), redirecting the edges incident to $n$ to the corresponding ports in the flattened Ladder graph: the function $\overline{Connect}_G$ in Algorithm~\ref{alg:alg5} is similar to $Connect_G$ except that if $e\in E_G$ was connected  to a port $p$ in a node $n$ with a ladder graph $W$, then $\overline{Connect}_G(e)$ returns the port corresponding to $p$ in the flattened version of $W$ (instead of $p$), which exists since the ladder graph $W$ has the same interface as $n$. In other words, $\mathcal{T}(G) = G'$ where $G'$ is obtained from $G$ by replacing each hierarchical node $n$ in $G$ with $\mathcal{T}(Ladder_G(n))$ and connecting edges incident to ports in $n$ to the corresponding ports in $\mathcal{T}(Ladder_G(n))$.

The recursive definition of the flattening function given in in Algorithm~\ref{alg:alg5} ensures the result is always a regular port graph (i.e, an AHP at level 0). Moreover, a hierarchical rewriting step induces a corresponding "flat" rewriting step (the converse does not hold since structural information gets lost in the flattening process).

\begin{property} Assuming there are no graph-type variables:

\begin{enumerate}
\item
If $G$ is an AHP, then $\mathcal{T}(G)$ is a regular attributed port graph. Similarly, the flattening of an AHP-rewrite rule produces a regular port graph rule.
\item
If $G$ is an AHP and $L \Ra_C R$ an AHP rewrite rule, such that
$G\Ra H$ using   $L \Ra_C R$,  then $\mathcal{T}(G) \Ra \mathcal{T}(H)$ using $\mathcal{T}(L \Ra_C R)$.
\end{enumerate}
\end{property}

\begin{proof} 
\begin{enumerate}
\item
The first part follows by induction: The base case is trivial: $G$ is an AHP at level 0, that is, a regular port graph (since there are no variables of type graph by assumption), and the translation returns the same graph.

If $G$ is an AHP at level $i$, then by induction (since the ladder graphs in $G$ are at a lower level) the set ${\cal W}$ computed by the flattening function contains regular attributed port graphs. Moreover, they are disjoint because by definition of AHP all the ladder graphs are disjoint.  The output of the flattening function is built using the port graphs in ${\cal W}$ and  the nodes in $G$ that do not contain ladders. To complete the proof we need to show that the functions $Connect_{G'}$, $Attach_{G'}$ and ${\cal L}_{G'}$ are well defined. 

We prove  first that the function $Connect_{G'}$ returns a pair of ports in $G'$ for each edge in $E_{G'}$:

All the edges in $G'$ are either in $E_G$ or in a graph $W \in {\cal W}$ (i.e. in the translation of a ladder graph). If $e \in E_G$, by definition $\overline{Connect}_G$ is defined and returns the ports where the edge is attached (if $e$ was connected in $G$ to a port $p$ in a node $n$ with a ladder graph, then $\overline{Connect}_G$ returns the port in $W$ corresponding to $p$, which exists since the ladder graph has the same interface as $n$). If $e \in W$ then $Connect_{G'}(e) = Connect_W(e)$ by definition of $\mathcal{T}$, and $Connect_W$ is well defined  by induction.

To show that  $Attach_{G'}$ returns a node in $V_{G'}$ for each port $p$ in $P_G'$, we observe that the ports in $P_{G'}$ are either in $P_G^0$ and therefore $Attach_G(p)$ is defined, or in one of the graphs $W$ obtained by flattening a ladder graph, in which case $Attach_W(p)$ is defined by induction. 

Finally, it is easy to see that ${\cal L}_G$ is well defined by induction and the assumption that $G$ is an AHP. 

Since AHP rules are AHP port graphs, their flattened version is a regular port graph. To show that it is indeed a port graph rewrite rule it is sufficient to notice that the arrow node does not have a Ladder graph, and therefore it is part of the flattened rule, together with the edges that link it to the left and right hand sides, as expected.
\item The second part is a consequence of the fact that the same flattening function is applied to the rule and to the graph that will be rewritten; if there is  an AHP matching morphism between the AHP $L$ and $G$, then there is a matching morphism 
between ${\cal T}(L)$ and ${\cal T}(G)$. 
\end{enumerate}
\end{proof}

\paragraph{SPO rewriting semantics:} L\"owe~\cite{LoweM:TCS} showed that if $Sig$ is an algebraic signature, all $Sig$-algebras and partial $Sig$-morphisms form a category $Alg^p(Sig)$. This category is not closed with respect to pushouts in general. However, if $Sig$ contains only unary operators, i.e., it is a \emph{graph structure} in L\"owe's terminology, then pushouts do exist (see~\cite{LoweM:TCS}, Theorem 2.7). This result also holds for attributed graph structures as shown in~\cite{Lowe:1993}. Attributed port graphs have been shown to be attributed graph structures in~\cite{grs_porgy_main_long}, by defining suitable signatures and interpreting an attributed port graph as an algebra with that signature. Here we show that AHPs are also attributed graph structures. The proof is more involved for AHPs due to the nesting of graphs. The key idea is to introduce a sort ``graph'' and operators to group nodes, and hence ports and edges into graphs at various levels.

First, we recall the definition of algebraic signature and graph structure and refer to~\cite{LoweM:TCS,Lowe:1993} for more details.  

\begin{definition}[Graph Structure]
\label{def:GS}
An algebraic signature $Sig = (S,Op)$ consists of a set $S$ of
\emph{sorts} and a set $Op$ of operator symbols. 

Given an algebraic signature $Sig = (S,Op)$, if $A,B$ are $Sig$-algebras, a partial $Sig$-morphism $h: A \mapsto B$ is
a total morphism from some sub-algebra $A_h$ of $A$ to $B$; $A_h$ is called the scope of $h$. 

A \emph{graph structure} is a signature that contains unary operators only.

If $GS=(S_1, OP_1)$ is a graph structure, $S$ a subset of $S_1$ and
$SIG=(S_2, OP_2)$ an arbitrary signature, a \emph{$SIG$-attribution} of $GS$
is an $S$-indexed family of operator symbols 
$ATTROP = (ATTROP_s : s \mapsto s2_s)_{s \in S}$ where $s \in S$ and
$s2 \in S_2$.

An \emph{attributed graph} is a $GS$-graph with attributes in $SIG$, i.e., an
algebra with respect to  the signature $ATTR = GS + SIG + ATTROP$.

A morphism $f\colon A \mapsto B$ between $GS$-graphs $A$ and $B$ having
attributes in $SIG$ is a partial $GS$-morphism $f1\colon (A)_{GS} \mapsto (B)_{GS}$ together with a total $SIG$-morphism
$f2 \colon (A)_{SIG} \mapsto (B)_{SIG}$ satisfying for all operators
$attr\colon s1 \mapsto s2 \in ATTROP$ and all $x \in A(f1)_{s1},
f2(attr^A(x)) = attr^B(f1(x))$.

A rewrite rule is an $ATTR$-morphism $r$ whose $SIG$-component is an
isomorphism $f2\colon (A)_{SIG} \mapsto (B)_{SIG}$.
\end{definition}

To show that AHP are attributed graph structures, we consider signatures $AH=(S,OP)$ and $SIG=(S_{1},OP_{1})$ such that:
\begin{itemize}
\item $S=\lbrace node,port,edge,graph,rec_{node}, rec_{port}, rec_{edge}, rec_{graph}, list[port]\rbrace$.
Formally, $list[port]$ is a family of sorts, one for each arity; we abbreviate it as one sort.
\item $OP$ is the set of operators including:
\begin{enumerate}
\item $s,t:edge \mapsto port$; these operators will be interpreted by the $Connect$ function.
\item $ports: node \mapsto list[port]$; this operator will be interpreted by using the $Attach$ function (again, formally it is a family of operators, one for each arity, but we abbreviate it as one operator).
\item $ladder_V: node \mapsto graph$; this operator will be interpreted as a function that returns the ladder graph to which the node belongs. 
\item $l_V: rec_{node} \mapsto node$, $l_P: rec_{port} \mapsto port$, $l_E: rec_{edge} \mapsto edge$, $l_{\cal G}: rec_{graph} \mapsto graph$; these operators will be interpreted as functions that return the node, port, edge or graph to which the record belongs.
\end{enumerate}
\item $S_1= D \cup \lbrace attribute,value,pair,record\rbrace$, where $D$ is the set of data sorts (for $values$), the other sorts are used to type the operators that build records.
\item $OP_1$ includes a set of operators on data sorts (such as arithmetic operators) and operators used to build records:
\begin{enumerate}
\item $\_ := \_: attribute, value \mapsto pair$
\item $\_.\_: record, attribute \mapsto value$
\item $\lbrace \_,\_\rbrace: pair, record \mapsto record$
\item $\lbrace \rbrace: \mapsto record$
\end{enumerate}
\end{itemize}
Let the $SIG$-attribution of $AH$, $ATTROP$, be such that $ATTROP_{rec_{node}}:rec_{node} \mapsto record$, $ATTROP_{rec_{port}}:rec_{port} \mapsto record$, and similarly for the other $rec$ sorts. We show below that an AHP $G$ can be seen as an algebra on the combination of three signatures $ATTR=AH+SIG+ATTROP$ and is therefore an attributed graph structure.

\begin{property}
AHP graphs are attributed graph structures.
\end{property}

\begin{proof}
Since all operators in $OP$ are unary, $AH$ is a  graph structure. To complete the proof we show that an AHP $G = (V, P, E, {\cal G},D)_\mathcal{F}$ can be seen as an algebra on the signature $ATTR=AH+SIG+ATTROP$:
\begin{itemize}
\item $V, P$, $E$, ${\cal G}$ are carriers of the sorts $node, port, edge, graph$ and the sort $rec_i$ ($i \in \{node,$ $port,$ $edge,$ $graph\})$ is interpreted by a set of pointers, one for each element of $V$, $P$, $E$, ${\cal G}$. 
\item $s$ and $t$ are interpreted by the function $Connect$: \\
$s,t:edge \mapsto port$ such that  $s(e):=p_1, t(e):=p_2$ iff $e \in E \wedge Connect(e)=(p_1,p_2)$.
\item $ports$ is interpreted by a function such that if $Attach(p_i) = n$ ($1\hspace*{1pt}{\leq}\hspace*{1pt} i\hspace*{1pt}{\leq}\hspace*{1pt} k$) then $ports(n)\hspace*{1pt}{:=}\hspace*{1pt} [p_1,\ldots, p_k]$.
\item $ladder_V$ 
is interpreted by a function such that if  $n$ is a node in a ladder graph $W$ ($n\in V_W$) then $ladder_V(n) =W$.
\item the injective operators $l_V$, $l_P$, $l_E$, $l_{\cal G}$ are interpreted using the respective AHP's labelling function $\mathcal{L}$ (for example, if $\mathcal{L}_G(e) = r$ then $l_E(r) = e$; the definitions of $l_V$ and $l_P$  are similar). 

\item The sub-algebra that corresponds to $SIG$ defines the interpretation for records (it can either be interpreted as a term-algebra with variables or a class of concrete objects, depending on whether we are considering a graph in a rewrite rule or a concrete graph to be rewritten).
\end{itemize}
This completes the proof.
\end{proof}

We can also show that AHP morphisms are ATTR morphisms. To complete the SPO semantics for rewriting, we need to show that AHP rules define a partial morphism from the left to the right-hand side. This was shown in \cite{grs_porgy_main_long} for \emph{simple} rules  where \emph{the arrow node maps one port in the left-hand side to only one port in the right-hand side}. Under the same condition, a similar result can be shown for AHP rules. The proof is more involved because there is an additional sort $graph$ and operators that map nodes to ladder graphs. However, since the partial morphism defined by the rule is generated 
by the map between ports in the top level of $L$ and $R$ (specified by the arrow node and its edges), no ladder graphs are involved (top level ports do not belong to ladder graphs, since all graph components in an AHP are disjoint).

\begin{property}
\begin{enumerate}
\item
A simple AHP rule is an $ATTR$-morphism $r$ whose $SIG$ component is an isomorphism, and a match $m$ between the left-hand side of a rule, $L$, and a redex in an AHP $G$ is an $ATTR$ morphism whose $AH$ component is total. 
\item
The application of a simple rule $r$ to a graph $G$ at redex $m(L)$ is the pushout of $r$ and $m$.
\end{enumerate}
\end{property}

\begin{proof}
To prove the first part, we first show that an AHP morphism between two AHP graphs (Definition~\ref{def:AHPmorphism}) 
maps to a partial morphism between $AH$-graphs $A$ and $B$ that have attributes in $SIG$, together with  a total $SIG$-morphism satisfying for all operators
$attr\colon s1 \mapsto s2 \in ATTROP$ and all $x \in A(f1)_{s1},
f2(attr^A(x)) = attr^B(f1(x))$:
\\In the AHP-morphism definition, a (partial) {\em morphism} $f$ from $G$ to $H$, denoted $f: G \rightarrow H$, with definition domain $Dom(f)$, is defined by a family of partial functions $\langle f_V:V_G \ra V_H, f_P: P_G \ra P_H, f_E:
E_G \ra E_H, f_{\cal G}: {\cal G}_G \mapsto {\cal G}_H, f_D: D_G \mapsto D_H\rangle$. This family of functions define a partial AH-morphism $f_1:(G)_{AH} \mapsto (H)_{AH}$ which coincides with $\langle f_V:V_G \ra V_H, f_P: P_G \ra P_H, f_E:
E_G \ra E_H, f_{\cal G}: {\cal G}_G \mapsto {\cal G}_H \rangle$ on the carriers of sorts \textit{node, port, edge, graph}.
The total SIG-morphism $f_2:(G)_{SIG} \mapsto (H)_{SIG}$ is the restriction of $f$ on records and coincides with $f_D:D_G \mapsto D_H$. $f_2$ satisfies $\forall attr:a'\mapsto a \in ATTROP$ and all $G(f_1)_s, f_2(attr^{G}(x)) = attr^H(f_1(x))$ due to the condition on $\mathcal{L}$ in definition~\ref{def:AHPmorphism} such that the morphism preserves record attributes and their values:\\
For all $n \in Dom(f)$, $f_{D}( {\cal L}_G(n) ) = {\cal L}_H (f_V(n))$  \\
For all $p \in Dom(f)$, $f_{D}( {\cal L}_G(p) )= {\cal L}_H (f_P(p))$ \\
For all $e \in Dom(f)$, $f_{D}( {\cal L}_G(e) ) = {\cal L}_H (f_E(e))$  \\
For all $W \in Dom(f)$, $f_{D}( {\cal L}_G(W) ) = {\cal L}_H (f_{\cal G}(W))$.

To show that a simple AHP rule defines a partial morphism between the left- and right-hand sides, notice that in a  simple rule  the arrow node links one port in $L$ to at most one port in $R$. Thus, the edges in the arrow node define a \emph{partial function} from $P_L$ (the set of ports in the left-hand side) to $P_R$ (the set of ports in the right-hand side). Since there are no operators on ports in the algebra, this is trivially a morphism.

Since records are implemented in the same way on both sides, an AHP rule can be seen as an ATTR-morphism whose SIG component is an isomorphism. This completes the proof of the first part.

To prove the second part, first notice that a match $m$ between the left-hand side of a rule, L, and a redex in an AHP $G$ is an ATTR-morphism whose AH-component is total. This follows from the definition of AHP morphism (Definition~\ref{def:AHPmorphism}).

The fact that a pushout of $r$ and $m$ is equivalent to the application of a simple rule $r$ to $G$ at redex $m(L)$ in order to produce $H$ is seen in the description of the steps involved in an AHP rewrite step $G \Ra H$, which map to the steps described in~\cite{Lowe:1993} to build the pushout object.

The gluing object consists of the ports in $L$ that are connected to the bridge ports. The pushout object is isomorphic to $G$ where $m(L)$ is replaced with $m(R)$ and the external edges connected to $m(dom(r))$ are redirected as indicated in the definition of rewriting step.
\end{proof}

\section{Related Work}
\label{section_related_work}
Various extensions of graph formalisms have been previously defined with the aim to provide abstraction and structuring features in graph-based modelling tools. A prominent example is the concept of  bigraph introduced by Milner (see~\cite{grs_bigraph_milner,BirkedalDGM07}) to model computation on a global scale. Bigraphs are graphs whose nodes may be nested, thus providing direct support to notions of locality (nesting of nodes) and connectivity (edges), which are key aspects of mobile systems~\cite{IandC2006}. Formally, bigraphs  are defined  in terms of two structures that share the same node set: place graphs and link graphs. A place graph is a forest (and in this sense bigraphs define a notion of hierarchy similar to AHP's), whereas the link graph is a hyper-graph. Closely related to bigraphs are the deduction graphs proposed by Geuvers and Loeb~\cite{DBLP:journals/mscs/GeuversL07} to represent proofs, and gs-graphs~\cite{DBLP:journals/fuin/BruniMPT14} can be proven to be essentially equivalent.  

Hierarchical hyper-graphs~\cite{grs_hgraph_hierarchical_pl_2000} also provide structuring features: attributes of type graph are permitted within hierarchical hyper-graphs in association with edges as ``frames'' where frames are hyper-edges that can contain hierarchical hyper-graphs of an arbitrary nesting depth  
and, also in a similar fashion to our solution,  edges cutting across components are prohibited. 

Similar to our proposal, a recursive flattening approach is highlighted, albeit via hyper-edge replacement as opposed to node replacement. A less restrictive version can be found in~\cite{DBLP:journals/jcss/Palacz04}.

Bigraphs and hierarchical hyper-graphs are equipped with a formal, categorical semantics. Where a double push-out graph transformation approach applies naturally to the transformation of hierarchical hyper-graphs, the dynamic theory of bigraphs relies on a notion of relative push-out. For AHP graphs, we follow the single push-out approach advocated in previous port graph transformation systems.  However, the notion of matching we define in this paper is purely operational, and can be easily implemented as an extension of existing port graph matching algorithms. 

H-Graphs~\cite{grs_hierarchical_benchmark_} model hierarchy by labelling nodes of set N, over a set of atoms A, or permitting an embedded directed sub-graph created from N within A. H-Graphs in practice are used to model run-time data structures for the definition of programming language semantics and H-Graph grammars model operations over these structures by substituting an atomic node with a H-Graph.
A benchmark developed in 2001~\cite{grs_hierarchical_benchmark_} analyses hierarchical graphs in terms of underlying graph structure, the nested packages and a coupling mechanism, investigating the two major approaches presented by H-graph grammars and hierarchical hyper-graphs, and providing a means by which to compare their properties uniformly. A similar framework can be found in~\cite{grs_hierarchical_packages_}.

Distributed hierarchical graphs form the basis upon which the agent-based OWL semantic web ontological model is built and in which graph transformations can take place at various levels of abstraction~\cite{doi:10.1504/IJDMB.2015.066334}. 
Other interesting hierarchical representations include that of L-Graphs in which edges and nodes are labelled by graphs~\cite{SCHNEIDER1993257}, $\mathcal{M}$-adhesive Graph Transformation Systems~\cite{Padberg2017} that provide a generic algebraic definition to cater for varying graphs types, that of the verbose property graphs used in graph analytics~\cite{JunghannsPR17}, and the implementation found in~\cite{SLUSARCZYK201795}; that also makes use of attributes and multi-typed sub-graphs. \cite{Padberg2017}~also contains details of multi-hierarchical graphs and graph groupings.

Specifically for port graphs, an abstract higher-order calculus inspired by the $\rho$-calculus~\cite{CirsteaKirchner01} is defined in~\cite{AndreiK08,AndreiK08c}, where terms can refer to
objects that are port-graphs and variables may range over rewrite rules. This calculus can be seen as a combination of first-order rewrite rules with abstraction, but the notion of graph morphism (which is the basis for matching and rewriting) is first-order. As already mentioned, the higher-order port graphs found in~\cite{grs_portgraph_hierarchical} include a class of nodes labelled by variables that can be instantiated by graphs, thus offering abstraction but no additional structuring capabilities.  

\section{Conclusions and Future Work}
\label{section_conclusion}
We have introduced AHP as a means to specify hierarchical models in a manner that is concise, visual, modular and feasible. Future work will include a full implementation of this design within \porgy, completing the construction of the case-studies outlined and developing a translation between HOPG and AHP formalisms. 
Our hierarchical constraint could be relaxed to permit edges linking nodes in graphs at different levels in the hierarchy, although this will make the implementation of matching more involved. Boundary-crossing edges are permitted  within some of the visual languages used in software modelling (e.g., to represent UML diagrams).
\bibliographystyle{eptcs}
\bibliography{Bibliography}

\begin{thebibliography}{10}
\providecommand{\bibitemdeclare}[2]{}
\providecommand{\surnamestart}{}
\providecommand{\surnameend}{}
\providecommand{\urlprefix}{Available at }
\providecommand{\url}[1]{\texttt{#1}}
\providecommand{\href}[2]{\texttt{#2}}
\providecommand{\urlalt}[2]{\href{#1}{#2}}
\providecommand{\doi}[1]{doi:\urlalt{http://dx.doi.org/#1}{#1}}
\providecommand{\bibinfo}[2]{#2}

\bibitemdeclare{article}{Accattoli15}
\bibitem{Accattoli15}
\bibinfo{author}{Beniamino \surnamestart Accattoli\surnameend}
  (\bibinfo{year}{2015}): \emph{\bibinfo{title}{Proof nets and the
  call-by-value {\(\lambda\)}-calculus}}.
\newblock {\sl \bibinfo{journal}{Theor. Comput. Sci.}} \bibinfo{volume}{606},
  pp. \bibinfo{pages}{2--24}, \doi{10.1016/j.tcs.2015.08.006}.

\bibitemdeclare{inproceedings}{grs_portgraph_hierarchical1}
\bibitem{grs_portgraph_hierarchical1}
\bibinfo{author}{Sandra \surnamestart Alves\surnameend},
  \bibinfo{author}{Maribel \surnamestart Fern{\'{a}}ndez\surnameend} \&
  \bibinfo{author}{Ian \surnamestart Mackie\surnameend} (\bibinfo{year}{2011}):
  \emph{\bibinfo{title}{A new graphical calculus of proofs}}.
\newblock In: {\sl \bibinfo{booktitle}{Proceedings 6th International Workshop
  on Computing with Terms and Graphs, {TERMGRAPH} 2011, Saarbr{\"{u}}cken,
  Germany, 2nd April 2011.}}, pp. \bibinfo{pages}{69--84},
  \doi{10.4204/EPTCS.48.8}.

\bibitemdeclare{article}{FIN_agents_absmodel_}
\bibitem{FIN_agents_absmodel_}
\bibinfo{author}{Kartik \surnamestart Anand\surnameend}, \bibinfo{author}{Alan
  \surnamestart Kirman\surnameend} \& \bibinfo{author}{Matteo \surnamestart
  Marsili\surnameend} (\bibinfo{year}{2013}): \emph{\bibinfo{title}{Epidemics
  of rules, rational negligence and market crashes}}.
\newblock {\sl \bibinfo{journal}{The European Journal of Finance}}
  \bibinfo{volume}{19}(\bibinfo{number}{5}), pp. \bibinfo{pages}{438--447},
  \doi{10.1080/1351847X.2011.601872}.

\bibitemdeclare{phdthesis}{Porgy_AndreiPhd}
\bibitem{Porgy_AndreiPhd}
\bibinfo{author}{Oana \surnamestart Andrei\surnameend} (\bibinfo{year}{2008}):
  \emph{\bibinfo{title}{A Rewriting Calculus for Graphs: Applications to
  Biology and Autonomous Systems. (Un calcul de r{\'{e}}{\'{e}}criture de
  graphes : applications {\`{a}} la biologie et aux syst{\`{e}}mes
  autonomes)}}.
\newblock Ph.D. thesis, \bibinfo{school}{National Polytechnic Institute of
  Lorraine, Nancy, France}.

\bibitemdeclare{inproceedings}{porgy_2011}
\bibitem{porgy_2011}
\bibinfo{author}{Oana \surnamestart Andrei\surnameend},
  \bibinfo{author}{Maribel \surnamestart Fern{\'a}ndez\surnameend},
  \bibinfo{author}{H{\'e}l{\`e}ne \surnamestart Kirchner\surnameend},
  \bibinfo{author}{Guy \surnamestart Melan\c{c}on\surnameend},
  \bibinfo{author}{Olivier \surnamestart Namet\surnameend} \&
  \bibinfo{author}{Bruno \surnamestart Pinaud\surnameend}
  (\bibinfo{year}{2011}): \emph{\bibinfo{title}{PORGY: Strategy-Driven
  Interactive Transformation of Graphs}}.
\newblock In: {\sl \bibinfo{booktitle}{TERMGRAPH}}, pp.
  \bibinfo{pages}{54--68}, \doi{10.4204/EPTCS.48.7}.

\bibitemdeclare{article}{AndreiK08}
\bibitem{AndreiK08}
\bibinfo{author}{Oana \surnamestart Andrei\surnameend} \&
  \bibinfo{author}{H{\'e}l{\`e}ne \surnamestart Kirchner\surnameend}
  (\bibinfo{year}{2008}): \emph{\bibinfo{title}{A Rewriting Calculus for
  Multigraphs with Ports}}.
\newblock {\sl \bibinfo{journal}{Electr. Notes Theor. Comput. Sci.}}
  \bibinfo{volume}{219}, pp. \bibinfo{pages}{67--82},
  \doi{10.1016/j.entcs.2008.10.035}.

\bibitemdeclare{incollection}{AndreiK08c}
\bibitem{AndreiK08c}
\bibinfo{author}{Oana \surnamestart Andrei\surnameend} \&
  \bibinfo{author}{Helene \surnamestart Kirchner\surnameend}
  (\bibinfo{year}{2009}): \emph{\bibinfo{title}{A Higher-Order Graph Calculus
  for Autonomic Computing}}.
\newblock In \bibinfo{editor}{Marina \surnamestart Lipshteyn\surnameend},
  \bibinfo{editor}{Vadim~E. \surnamestart Levit\surnameend} \&
  \bibinfo{editor}{Ross~M. \surnamestart {McConnell}\surnameend}, editors: {\sl
  \bibinfo{booktitle}{Graph Theory, Computational Intelligence and Thought}},
  \bibinfo{publisher}{Springer-Verlag}, \bibinfo{address}{Berlin, Heidelberg},
  pp. \bibinfo{pages}{15--26}, \doi{10.1007/978-3-642-02029-2\_2}.

\bibitemdeclare{incollection}{DBLP:books/el/leeuwen90/Barendregt90}
\bibitem{DBLP:books/el/leeuwen90/Barendregt90}
\bibinfo{author}{Hendrik~Pieter \surnamestart Barendregt\surnameend}
  (\bibinfo{year}{1990}): \emph{\bibinfo{title}{Functional Programming and
  Lambda Calculus}}.
\newblock In: {\sl \bibinfo{booktitle}{Handbook of Theoretical Computer
  Science, Volume {B:} Formal Models and Sematics {(B)}}}, pp.
  \bibinfo{pages}{321--363}, \doi{10.1016/B978-0-444-88074-1.50012-3}.

\bibitemdeclare{article}{BirkedalDGM07}
\bibitem{BirkedalDGM07}
\bibinfo{author}{Lars \surnamestart Birkedal\surnameend},
  \bibinfo{author}{Troels~Christoffer \surnamestart Damgaard\surnameend},
  \bibinfo{author}{Arne~J. \surnamestart Glenstrup\surnameend} \&
  \bibinfo{author}{Robin \surnamestart Milner\surnameend}
  (\bibinfo{year}{2007}): \emph{\bibinfo{title}{Matching of Bigraphs}}.
\newblock {\sl \bibinfo{journal}{Electr. Notes Theor. Comput. Sci.}}
  \bibinfo{volume}{175}(\bibinfo{number}{4}), pp. \bibinfo{pages}{3--19},
  \doi{10.1016/j.entcs.2007.04.013}.

\bibitemdeclare{article}{DBLP:journals/fuin/BruniMPT14}
\bibitem{DBLP:journals/fuin/BruniMPT14}
\bibinfo{author}{Roberto \surnamestart Bruni\surnameend}, \bibinfo{author}{Ugo
  \surnamestart Montanari\surnameend}, \bibinfo{author}{Gordon~D. \surnamestart
  Plotkin\surnameend} \& \bibinfo{author}{Daniele \surnamestart
  Terreni\surnameend} (\bibinfo{year}{2014}): \emph{\bibinfo{title}{On
  Hierarchical Graphs: Reconciling Bigraphs, Gs-monoidal Theories and
  Gs-graphs}}.
\newblock {\sl \bibinfo{journal}{Fundam. Inform.}}
  \bibinfo{volume}{134}(\bibinfo{number}{3-4}), pp. \bibinfo{pages}{287--317},
  \doi{10.3233/FI-2014-1103}.

\bibitemdeclare{article}{grs_hierarchical_benchmark_}
\bibitem{grs_hierarchical_benchmark_}
\bibinfo{author}{Giorgio \surnamestart Busatto\surnameend} \&
  \bibinfo{author}{Berthold \surnamestart Hoffmann\surnameend}
  (\bibinfo{year}{2001}): \emph{\bibinfo{title}{Comparing Notions of
  Hierarchical Graph Transformation}}.
\newblock {\sl \bibinfo{journal}{Electr. Notes Theor. Comput. Sci.}}
  \bibinfo{volume}{50}(\bibinfo{number}{3}), pp. \bibinfo{pages}{310--317},
  \doi{10.1016/S1571-0661(04)00184-7}.

\bibitemdeclare{article}{grs_hierarchical_packages_}
\bibitem{grs_hierarchical_packages_}
\bibinfo{author}{Giorgio \surnamestart Busatto\surnameend},
  \bibinfo{author}{Hans{-}J{\"{o}}rg \surnamestart Kreowski\surnameend} \&
  \bibinfo{author}{Sabine \surnamestart Kuske\surnameend}
  (\bibinfo{year}{2005}): \emph{\bibinfo{title}{Abstract hierarchical graph
  transformation}}.
\newblock {\sl \bibinfo{journal}{Mathematical Structures in Computer Science}}
  \bibinfo{volume}{15}(\bibinfo{number}{4}), pp. \bibinfo{pages}{773--819},
  \doi{10.1017/S0960129505004846}.

\bibitemdeclare{article}{CirsteaKirchner01}
\bibitem{CirsteaKirchner01}
\bibinfo{author}{Horatiu \surnamestart Cirstea\surnameend} \&
  \bibinfo{author}{Claude \surnamestart Kirchner\surnameend}
  (\bibinfo{year}{2001}): \emph{\bibinfo{title}{The rewriting calculus ---
  {Part~I {\em and} II}}}.
\newblock {\sl \bibinfo{journal}{Logic Journal of the Interest Group in Pure
  and Applied Logics}} \bibinfo{volume}{9}(\bibinfo{number}{3}), pp.
  \bibinfo{pages}{427--498}, \doi{10.1093/jigpal/9.3.339}.

\bibitemdeclare{inproceedings}{Corradini:handbook}
\bibitem{Corradini:handbook}
\bibinfo{author}{Andrea \surnamestart Corradini\surnameend},
  \bibinfo{author}{Ugo \surnamestart Montanari\surnameend},
  \bibinfo{author}{Francesca \surnamestart Rossi\surnameend},
  \bibinfo{author}{Hartmut \surnamestart Ehrig\surnameend},
  \bibinfo{author}{Reiko \surnamestart Heckel\surnameend} \&
  \bibinfo{author}{Michael \surnamestart L{\"o}we\surnameend}
  (\bibinfo{year}{1997}): \emph{\bibinfo{title}{Algebraic Approaches to Graph
  Transformation - Part I: Basic Concepts and Double Pushout Approach}}.
\newblock In: {\sl \bibinfo{booktitle}{Handbook of Graph Grammars and Computing
  by Graph Transformations, Volume 1: Foundations}}, \bibinfo{publisher}{World
  Scientific}, pp. \bibinfo{pages}{163--246},
  \doi{10.1142/9789812384720\_0003}.
\newblock
  \urlprefix\url{http://www.worldscientific.com/doi/abs/10.1142/9789812384720\_0003}.

\bibitemdeclare{misc}{grs_stochastic_kappa_only_refinement_rules_}
\bibitem{grs_stochastic_kappa_only_refinement_rules_}
\bibinfo{author}{Vincent \surnamestart Danos\surnameend},
  \bibinfo{author}{\surnamestart Feret\surnameend}, \bibinfo{author}{Walter
  \surnamestart Fontana\surnameend}, \bibinfo{author}{Russell \surnamestart
  Harmer\surnameend}, \bibinfo{author}{Jonathan \surnamestart
  Hayman\surnameend}, \bibinfo{author}{Jean \surnamestart Krivine\surnameend},
  \bibinfo{author}{Chris \surnamestart Thompson-walsh\surnameend} \&
  \bibinfo{author}{Glynn \surnamestart Winskel\surnameend}:
  \emph{\bibinfo{title}{Graphs, rewriting and causality in rule-based models}},
  \doi{http://citeseerx.ist.psu.edu/viewdoc/summary?doi=10.1.1.221.6822}.

\bibitemdeclare{article}{grs_hgraph_hierarchical_pl_2000}
\bibitem{grs_hgraph_hierarchical_pl_2000}
\bibinfo{author}{Frank \surnamestart Drewes\surnameend},
  \bibinfo{author}{Berthold \surnamestart Hoffmann\surnameend} \&
  \bibinfo{author}{Detlef \surnamestart Plump\surnameend}
  (\bibinfo{year}{2002}): \emph{\bibinfo{title}{Hierarchical Graph
  Transformation}}.
\newblock {\sl \bibinfo{journal}{Journal of Computer and System Sciences}}
  \bibinfo{volume}{64}(\bibinfo{number}{2}), pp. \bibinfo{pages}{249 -- 283},
  \doi{10.1006/jcss.2001.1790}.

\bibitemdeclare{article}{DBLP:journals/entcs/EetveldeJ03}
\bibitem{DBLP:journals/entcs/EetveldeJ03}
\bibinfo{author}{Niels~Van \surnamestart Eetvelde\surnameend} \&
  \bibinfo{author}{Dirk \surnamestart Janssens\surnameend}
  (\bibinfo{year}{2003}): \emph{\bibinfo{title}{A Hierarchical Program
  Representation for Refactoring}}.
\newblock {\sl \bibinfo{journal}{Electr. Notes Theor. Comput. Sci.}}
  \bibinfo{volume}{82}(\bibinfo{number}{7}), pp. \bibinfo{pages}{91--104},
  \doi{10.1016/S1571-0661(04)80749-7}.

\bibitemdeclare{incollection}{DBLP:conf/gg/EhrigHKLRWC97}
\bibitem{DBLP:conf/gg/EhrigHKLRWC97}
\bibinfo{author}{Hartmut \surnamestart Ehrig\surnameend},
  \bibinfo{author}{Reiko \surnamestart Heckel\surnameend},
  \bibinfo{author}{Martin \surnamestart Korff\surnameend},
  \bibinfo{author}{Michael \surnamestart L{\"{o}}we\surnameend},
  \bibinfo{author}{Leila \surnamestart Ribeiro\surnameend},
  \bibinfo{author}{Annika \surnamestart Wagner\surnameend} \&
  \bibinfo{author}{Andrea \surnamestart Corradini\surnameend}
  (\bibinfo{year}{1997}): \emph{\bibinfo{title}{Algebraic Approaches to Graph
  Transformation - Part {II:} Single Pushout Approach and Comparison with
  Double Pushout Approach}}.
\newblock In: {\sl \bibinfo{booktitle}{Handbook of Graph Grammars and Computing
  by Graph Transformations, Volume 1: Foundations}}, pp.
  \bibinfo{pages}{247--312}, \doi{10.1142/9789812384720\_0004}.

\bibitemdeclare{misc}{WADT_Paper_}
\bibitem{WADT_Paper_}
\bibinfo{author}{Nneka \surnamestart Ene\surnameend}, \bibinfo{author}{Maribel
  \surnamestart Fern\'andez\surnameend} \& \bibinfo{author}{Bruno \surnamestart
  Pinaud\surnameend}: \emph{\bibinfo{title}{Graph Models for Capital Markets}},
  \doi{https://nms.kcl.ac.uk/nneka.ene/papers.html}.
\newblock \bibinfo{note}{Available from
  https://nms.kcl.ac.uk/nneka.ene/papers.html}.

\bibitemdeclare{inbook}{Engels2000}
\bibitem{Engels2000}
\bibinfo{author}{Gregor \surnamestart Engels\surnameend} \&
  \bibinfo{author}{Reiko \surnamestart Heckel\surnameend}
  (\bibinfo{year}{2000}): \emph{\bibinfo{title}{Graph Transformation as a
  Conceptual and Formal Framework for System Modeling and Model Evolution}},
  pp. \bibinfo{pages}{127--150}.
\newblock \bibinfo{publisher}{Springer Berlin Heidelberg},
  \bibinfo{address}{Berlin, Heidelberg}, \doi{10.1007/3-540-45022-X\_12}.

\bibitemdeclare{techreport}{grs_porgy_main_long}
\bibitem{grs_porgy_main_long}
\bibinfo{author}{Maribel \surnamestart Fern{\'a}ndez\surnameend},
  \bibinfo{author}{H{\'e}l{\`e}ne \surnamestart Kirchner\surnameend} \&
  \bibinfo{author}{Bruno \surnamestart Pinaud\surnameend}
  (\bibinfo{year}{2017}): \emph{\bibinfo{title}{{Strategic Port Graph
  Rewriting: an Interactive Modelling Framework}}}.
\newblock \bibinfo{type}{Research Report}, \bibinfo{institution}{{Inria ; LaBRI
  - Laboratoire Bordelais de Recherche en Informatique ; King's College
  London}}, \doi{https://hal.inria.fr/hal-01251871}.
\newblock \urlprefix\url{https://hal.inria.fr/hal-01251871}.

\bibitemdeclare{article}{FernandezMP07}
\bibitem{FernandezMP07}
\bibinfo{author}{Maribel \surnamestart Fern{\'{a}}ndez\surnameend},
  \bibinfo{author}{Ian \surnamestart Mackie\surnameend} \&
  \bibinfo{author}{Jorge~Sousa \surnamestart Pinto\surnameend}
  (\bibinfo{year}{2007}): \emph{\bibinfo{title}{A Higher-Order Calculus for
  Graph Transformation}}.
\newblock {\sl \bibinfo{journal}{Electr. Notes Theor. Comput. Sci.}}
  \bibinfo{volume}{72}(\bibinfo{number}{1}), pp. \bibinfo{pages}{45--58},
  \doi{10.1016/j.entcs.2002.09.005}.

\bibitemdeclare{inproceedings}{grs_portgraph_hierarchical}
\bibitem{grs_portgraph_hierarchical}
\bibinfo{author}{Maribel \surnamestart Fern{\'{a}}ndez\surnameend} \&
  \bibinfo{author}{S{\'{e}}bastien \surnamestart Maulat\surnameend}
  (\bibinfo{year}{2012}): \emph{\bibinfo{title}{Higher-order port-graph
  rewriting}}.
\newblock In: {\sl \bibinfo{booktitle}{Proceedings 2nd International Workshop
  on Linearity, {LINEARITY} 2012, Tallinn, Estonia, 1 April 2012.}}, pp.
  \bibinfo{pages}{25--37}, \doi{10.4204/EPTCS.101.3}.

\bibitemdeclare{article}{DBLP:journals/mscs/GeuversL07}
\bibitem{DBLP:journals/mscs/GeuversL07}
\bibinfo{author}{Herman \surnamestart Geuvers\surnameend} \&
  \bibinfo{author}{Iris \surnamestart Loeb\surnameend} (\bibinfo{year}{2007}):
  \emph{\bibinfo{title}{Natural deduction via graphs: formal definition and
  computation rules}}.
\newblock {\sl \bibinfo{journal}{Mathematical Structures in Computer Science}}
  \bibinfo{volume}{17}(\bibinfo{number}{3}), pp. \bibinfo{pages}{485--526},
  \doi{10.1017/S0960129507006123}.

\bibitemdeclare{inproceedings}{DBLP:conf/popl/GonthierAL92}
\bibitem{DBLP:conf/popl/GonthierAL92}
\bibinfo{author}{Georges \surnamestart Gonthier\surnameend},
  \bibinfo{author}{Mart{\'{\i}}n \surnamestart Abadi\surnameend} \&
  \bibinfo{author}{Jean{-}Jacques \surnamestart L{\'{e}}vy\surnameend}
  (\bibinfo{year}{1992}): \emph{\bibinfo{title}{The Geometry of Optimal Lambda
  Reduction}}.
\newblock In: {\sl \bibinfo{booktitle}{Conference Record of the Nineteenth
  Annual {ACM} {SIGPLAN-SIGACT} Symposium on Principles of Programming
  Languages, Albuquerque, New Mexico, USA, January 19-22, 1992}}, pp.
  \bibinfo{pages}{15--26}, \doi{10.1145/143165.143172}.

\bibitemdeclare{techreport}{fin_sec_model}
\bibitem{fin_sec_model}
\bibinfo{author}{Gary \surnamestart Gorton\surnameend} \&
  \bibinfo{author}{Andrew \surnamestart Metrick\surnameend}
  (\bibinfo{year}{2012}): \emph{\bibinfo{title}{Securitization}}.
\newblock \bibinfo{type}{Working Paper} \bibinfo{number}{18611},
  \bibinfo{institution}{National Bureau of Economic Research},
  \doi{http://www.nber.org/papers/w18611}.
\newblock \urlprefix\url{http://www.nber.org/papers/w18611}.

\bibitemdeclare{article}{HabelMP01}
\bibitem{HabelMP01}
\bibinfo{author}{Annegret \surnamestart Habel\surnameend},
  \bibinfo{author}{J{\"u}rgen \surnamestart M{\"u}ller\surnameend} \&
  \bibinfo{author}{Detlef \surnamestart Plump\surnameend}
  (\bibinfo{year}{2001}): \emph{\bibinfo{title}{Double-pushout graph
  transformation revisited}}.
\newblock {\sl \bibinfo{journal}{Mathematical Structures in Computer Science}}
  \bibinfo{volume}{11}(\bibinfo{number}{5}), pp. \bibinfo{pages}{637--688},
  \doi{10.1017/S0960129501003425}.

\bibitemdeclare{inproceedings}{JunghannsPR17}
\bibitem{JunghannsPR17}
\bibinfo{author}{Martin \surnamestart Junghanns\surnameend},
  \bibinfo{author}{Andr{\'{e}} \surnamestart Petermann\surnameend} \&
  \bibinfo{author}{Erhard \surnamestart Rahm\surnameend}
  (\bibinfo{year}{2017}): \emph{\bibinfo{title}{Distributed Grouping of
  Property Graphs with Gradoop}}.
\newblock In: {\sl \bibinfo{booktitle}{Proc. Datenbanksysteme für Business,
  Technologie und Web (BTW)}}, pp. \bibinfo{pages}{103--122}.

\bibitemdeclare{inproceedings}{DBLP:conf/popl/Lafont90}
\bibitem{DBLP:conf/popl/Lafont90}
\bibinfo{author}{Yves \surnamestart Lafont\surnameend} (\bibinfo{year}{1990}):
  \emph{\bibinfo{title}{Interaction Nets}}.
\newblock In: {\sl \bibinfo{booktitle}{Conference Record of the Seventeenth
  Annual {ACM} Symposium on Principles of Programming Languages, San Francisco,
  California, USA, January 1990}}, pp. \bibinfo{pages}{95--108},
  \doi{10.1145/96709.96718}.

\bibitemdeclare{article}{LoweM:TCS}
\bibitem{LoweM:TCS}
\bibinfo{author}{Michael \surnamestart L\"owe\surnameend}
  (\bibinfo{year}{1993}): \emph{\bibinfo{title}{Algebraic approach to
  single-pushout graph transformation}}.
\newblock {\sl \bibinfo{journal}{Theoretical Computer Science}}
  \bibinfo{volume}{109}, pp. \bibinfo{pages}{181--224},
  \doi{10.1016/0304-3975(93)90068-5}.

\bibitemdeclare{incollection}{Lowe:1993}
\bibitem{Lowe:1993}
\bibinfo{author}{Michael \surnamestart L\"{o}we\surnameend},
  \bibinfo{author}{Martin \surnamestart Korff\surnameend} \&
  \bibinfo{author}{Annika \surnamestart Wagner\surnameend}
  (\bibinfo{year}{1993}): \emph{\bibinfo{title}{An Algebraic Framework for the
  Transformation of Attributed Graphs}}.
\newblock In \bibinfo{editor}{M.~R. \surnamestart Sleep\surnameend},
  \bibinfo{editor}{M.~J. \surnamestart Plasmeijer\surnameend} \&
  \bibinfo{editor}{M.~C. J.~D. \surnamestart van Eekelen\surnameend}, editors:
  {\sl \bibinfo{booktitle}{Term Graph Rewriting}}, \bibinfo{publisher}{John
  Wiley and Sons Ltd.}, \bibinfo{address}{Chichester, UK}, pp.
  \bibinfo{pages}{185--199}.

\bibitemdeclare{article}{RePEc:pal:jbkreg:v:14:y:2013:i:3:p:285-305}
\bibitem{RePEc:pal:jbkreg:v:14:y:2013:i:3:p:285-305}
\bibinfo{author}{Sheri \surnamestart Markose\surnameend}
  (\bibinfo{year}{2013}): \emph{\bibinfo{title}{Systemic risk analytics: A
  data-driven multi-agent financial network (MAFN) approach}}.
\newblock {\sl \bibinfo{journal}{Journal of Banking Regulation}}
  \bibinfo{volume}{14}(\bibinfo{number}{3-4}), pp. \bibinfo{pages}{285--305},
  \doi{10.1057/jbr.2013.10}.

\bibitemdeclare{misc}{FIN_ABM_Essex_CRTransferModel_}
\bibitem{FIN_ABM_Essex_CRTransferModel_}
\bibinfo{author}{Sheri \surnamestart Markose\surnameend}, \bibinfo{author}{Yang
  \surnamestart Dong\surnameend} \& \bibinfo{author}{Bewaji \surnamestart
  Oluwasegun\surnameend} (\bibinfo{year}{2008}): \emph{\bibinfo{title}{An
  Multi-Agent Model of RMBS, Credit Risk Transfer in Banks and Financial
  Stability: Implications of the Subprime Crisis}},
  \doi{http://citeseerx.ist.psu.edu/viewdoc/download?doi=10.1.1.509.3202\&rep=rep1\&type=pdf}.

\bibitemdeclare{incollection}{bath48096}
\bibitem{bath48096}
\bibinfo{author}{Sheri~M. \surnamestart Markose\surnameend},
  \bibinfo{author}{Bewaji \surnamestart Oluwasegun\surnameend} \&
  \bibinfo{author}{Simone \surnamestart Giansante\surnameend}
  (\bibinfo{year}{2014}): \emph{\bibinfo{title}{Multi-agent financial network
  (MAFN) model of {US} collateralized debt obligations ({CDO}):regulatory
  capital arbitrage, negative {CDS} carry trade, and systemic risk analysis}}.
\newblock In: {\sl \bibinfo{booktitle}{Banking, Finance, and Accounting}},
  \bibinfo{publisher}{IGI Global}, \bibinfo{address}{Hershey, {U. S. A.}}, pp.
  \bibinfo{pages}{561--590}, \doi{10.4018/978-1-4666-6268-1.ch030}.

\bibitemdeclare{article}{2006q.bio.....4006M}
\bibitem{2006q.bio.....4006M}
\bibinfo{author}{O.~\surnamestart {Mason}\surnameend} \&
  \bibinfo{author}{M.~\surnamestart {Verwoerd}\surnameend}:
  \emph{\bibinfo{title}{{Graph Theory and Networks in Biology}}}.
\newblock {\sl \bibinfo{journal}{IET Systems Biology}}
  \bibinfo{volume}{1}(\bibinfo{number}{2}), pp. \bibinfo{pages}{89--119},
  \doi{10.1049/iet-syb:20060038}.

\bibitemdeclare{techreport}{grs_bigraph_milner}
\bibitem{grs_bigraph_milner}
\bibinfo{author}{Robin \surnamestart Milner\surnameend} (\bibinfo{year}{2001}):
  \emph{\bibinfo{title}{Bigraphical reactive systems: basic theory}}.
\newblock \bibinfo{type}{Technical Report} \bibinfo{number}{UCAM-CL-TR-523},
  \bibinfo{institution}{University of Cambridge, Computer Laboratory}.

\bibitemdeclare{article}{IandC2006}
\bibitem{IandC2006}
\bibinfo{author}{Robin \surnamestart Milner\surnameend} (\bibinfo{year}{2006}):
  \emph{\bibinfo{title}{Pure bigraphs: Structure and dynamics}}.
\newblock {\sl \bibinfo{journal}{Inf. Comput.}}
  \bibinfo{volume}{204}(\bibinfo{number}{1}), pp. \bibinfo{pages}{60--122},
  \doi{10.1016/j.ic.2005.07.003}.

\bibitemdeclare{inproceedings}{Padberg2017}
\bibitem{Padberg2017}
\bibinfo{author}{Julia \surnamestart Padberg\surnameend}
  (\bibinfo{year}{2017}): \emph{\bibinfo{title}{Hierarchical Graph
  Transformation Revisited - Transformations of Coalgebraic Graphs}}.
\newblock In: {\sl \bibinfo{booktitle}{Graph Transformation - 10th
  International Conference, {ICGT} 2017, Held as Part of {STAF} 2017, Marburg,
  Germany, July 18-19, 2017, Proceedings}}, pp. \bibinfo{pages}{20--35},
  \doi{10.1007/978-3-319-61470-0\_2}.

\bibitemdeclare{article}{DBLP:journals/jcss/Palacz04}
\bibitem{DBLP:journals/jcss/Palacz04}
\bibinfo{author}{Wojciech \surnamestart Palacz\surnameend}
  (\bibinfo{year}{2004}): \emph{\bibinfo{title}{Algebraic hierarchical graph
  transformation}}.
\newblock {\sl \bibinfo{journal}{J. Comput. Syst. Sci.}}
  \bibinfo{volume}{68}(\bibinfo{number}{3}), pp. \bibinfo{pages}{497--520},
  \doi{10.1016/S0022-0000(03)00064-3}.

\bibitemdeclare{incollection}{Plump98termgraph}
\bibitem{Plump98termgraph}
\bibinfo{author}{Detlef \surnamestart Plump\surnameend} (\bibinfo{year}{1998}):
  \emph{\bibinfo{title}{Term Graph Rewriting}}.
\newblock In \bibinfo{editor}{Hartmut \surnamestart Ehrig\surnameend},
  \bibinfo{editor}{Gregor \surnamestart Engels\surnameend},
  \bibinfo{editor}{Hans-J{\"o}rg \surnamestart Kreowski\surnameend} \&
  \bibinfo{editor}{Grzegorz \surnamestart Rozenberg\surnameend}, editors: {\sl
  \bibinfo{booktitle}{Handbook of Graph Grammars and Computing by Graph
  Transformations, Volume 2: Applications, Languages, and Tools}},
  \bibinfo{publisher}{World Scientific}, pp. \bibinfo{pages}{3--61},
  \doi{10.1142/9789812815149\_0001}.
\newblock
  \urlprefix\url{http://www.worldscientific.com/doi/abs/10.1142/9789812815149\_0001}.

\bibitemdeclare{article}{SCHNEIDER1993257}
\bibitem{SCHNEIDER1993257}
\bibinfo{author}{H.J. \surnamestart Schneider\surnameend}
  (\bibinfo{year}{1993}): \emph{\bibinfo{title}{On categorical graph grammars
  integrating structural transformations and operations on labels}}.
\newblock {\sl \bibinfo{journal}{Theoretical Computer Science}}
  \bibinfo{volume}{109}(\bibinfo{number}{1}), pp. \bibinfo{pages}{257 -- 274},
  \doi{10.1016/0304-3975(93)90070-A}.

\bibitemdeclare{article}{doi:10.1504/IJDMB.2015.066334}
\bibitem{doi:10.1504/IJDMB.2015.066334}
\bibinfo{author}{Arash \surnamestart Shaban–Nejad\surnameend} \&
  \bibinfo{author}{Volker \surnamestart Haarslev\surnameend}
  (\bibinfo{year}{2015}): \emph{\bibinfo{title}{Managing changes in distributed
  biomedical ontologies using hierarchical distributed graph transformation}}.
\newblock {\sl \bibinfo{journal}{International Journal of Data Mining and
  Bioinformatics}} \bibinfo{volume}{11}(\bibinfo{number}{1}), pp.
  \bibinfo{pages}{53--83}, \doi{10.1504/IJDMB.2015.066334}.

\bibitemdeclare{article}{RuleBender}
\bibitem{RuleBender}
\bibinfo{author}{Adam~M. \surnamestart Smith\surnameend}, \bibinfo{author}{Wen
  \surnamestart Xu\surnameend}, \bibinfo{author}{Yao \surnamestart
  Sun\surnameend}, \bibinfo{author}{James~R. \surnamestart Faeder\surnameend}
  \& \bibinfo{author}{G.Elisabeta \surnamestart Marai\surnameend}
  (\bibinfo{year}{2012}): \emph{\bibinfo{title}{RuleBender: integrated
  modeling, simulation and visualization for rule-based intracellular
  biochemistry}}.
\newblock {\sl \bibinfo{journal}{BMC Bioinformatics}}
  \bibinfo{volume}{13}(\bibinfo{number}{8}):\bibinfo{eid}{S3},
  \doi{10.1186/1471-2105-13-S8-S3}.

\bibitemdeclare{inproceedings}{ValletKPM15}
\bibitem{ValletKPM15}
\bibinfo{author}{Jason \surnamestart Vallet\surnameend},
  \bibinfo{author}{H{\'{e}}l{\`{e}}ne \surnamestart Kirchner\surnameend},
  \bibinfo{author}{Bruno \surnamestart Pinaud\surnameend} \&
  \bibinfo{author}{Guy \surnamestart Melan{\c{c}}on\surnameend}
  (\bibinfo{year}{2015}): \emph{\bibinfo{title}{A Visual Analytics Approach to
  Compare Propagation Models in Social Networks}}.
\newblock In: {\sl \bibinfo{booktitle}{Proc. Graphs as Models, GaM 2015}}, pp.
  \bibinfo{pages}{65--79}, \doi{10.4204/EPTCS.181.5}.

\bibitemdeclare{article}{SLUSARCZYK201795}
\bibitem{SLUSARCZYK201795}
\bibinfo{author}{Grażyna \surnamestart Ślusarczyk\surnameend},
  \bibinfo{author}{Andrzej \surnamestart Łachwa\surnameend},
  \bibinfo{author}{Wojciech \surnamestart Palacz\surnameend},
  \bibinfo{author}{Barbara \surnamestart Strug\surnameend},
  \bibinfo{author}{Anna \surnamestart Paszyńska\surnameend} \&
  \bibinfo{author}{Ewa \surnamestart Grabska\surnameend}
  (\bibinfo{year}{2017}): \emph{\bibinfo{title}{An extended hierarchical
  graph-based building model for design and engineering problems}}.
\newblock {\sl \bibinfo{journal}{Automation in Construction}}
  \bibinfo{volume}{74}, pp. \bibinfo{pages}{95 -- 102},
  \doi{10.1016/j.autcon.2016.11.008}.

\end{thebibliography}
\end{document}